\documentclass{eptcs}
\usepackage{amssymb}
\usepackage{graphicx}
\usepackage[T1]{fontenc}
\usepackage{url}
\usepackage{color}
 \usepackage{booktabs}

\title{Formal Reasoning Using an Iterative Approach with an Integrated Web IDE}
\author{Nabil M. Kabbani, Daniel Welch, Caleb Priester, Stephen Schaub,\\ Blair Durkee, Yu-Shan Sun, and Murali Sitaraman
\institute{Clemson University,\\ Clemson SC 29631, USA}
\email{\{nkabban,dtwelch,cpriest,sschaub,bdurkee,yushans,msitara\}@clemson.edu}
}

\def\titlerunning{Formal Reasoning Using an Integrated Web IDE}
\def\authorrunning{Nabil M. Kabbani, Daniel Welch, et al.}

\begin{document}
\maketitle

\begin{abstract}
This paper summarizes our experience in communicating the elements of reasoning about correctness, and the central role of formal specifications in reasoning about modular, component-based software using a language and an integrated Web IDE designed for the purpose. Our experience in using such an IDE, supported by a `push-button' verifying compiler in a classroom setting, reveals the highly iterative process learners use to arrive at suitably specified, automatically provable code. We explain how the IDE facilitates reasoning at each step of this process by providing human readable verification conditions (VCs) and feedback from an integrated prover that clearly indicates unprovable VCs to help identify obstacles to completing proofs. The paper discusses the IDE's usage in verified software development using several examples drawn from actual classroom lectures and student assignments to illustrate principles of design-by-contract and the iterative process of creating and subsequently refining assertions, such as loop invariants in object-based code.  
\end{abstract}

\section{Introduction}

An IDE equipped with a formal verification system at its back end can facilitate an alternative style of developing software that involves using feedback from the verifier to locate and correct errors statically, instead of a more classical testing and debugging approach.  This paper illustrates how such an approach can work in practice based on our experience in employing it to teach a software engineering course, where students are asked to develop software components that are provably correct with respect to a set of given specifications.

The discussion in this paper is in the context of teaching analytical reasoning to undergraduate CS students.  The overall details of our educational approach for teaching mathematical reasoning, including an evaluation of student learning over multiple years in two required courses at Clemson, may be found in \cite{drachova:2013,drachova:2015}. Details of the types of software component development and reasoning assignments given to students are the topic of \cite{cook:2013}. The purpose of the present paper is to explain the iterative approach that students and software engineers, in general, can employ for developing verifiably correct software using the feedback from the Web IDE and its integrated prover.  

The IDE discussed in this paper is web-integrated, easy to use, and freely available online. It has been used at multiple institutions over the span of several years for teaching \cite{cook:2014} and research \cite{welch:2014} purposes, and is designed for RESOLVE, an integrated specification and programming language supported by a push-button verifying compiler \cite{sitaraman:2011}. The characteristics that distinguish the RESOLVE language and approach from most others include the following \cite{kulczycki:2004}: 

\begin{itemize}
\item Mathematical theories used in specifying programming concepts are extensible and they are described in reusable mathematical units; The theories are purely mathematical and do not involve any computational considerations.  They are carefully engineered to ease automated proving.  Number theory and a theory of strings over arbitrary types (used in this paper) are some examples.

\item Specifications of programming concepts that encapsulate abstract data types are kept strictly separate from implementations to facilitate design-by-contract \cite{meyer:1997} and to allow for multiple ways of realizing the same concept and permit efficiency trade-offs.  Examples of such concepts include programming integers with computational bounds, arrays, stacks, queues, and lists. 

\item The notion of clean semantics \cite{kulczycki:2004} makes it inessential to introduce and reason about object references explicitly in typical user code.  
\end{itemize}

While the literature includes several integrated development environments based on formal techniques related to ours (see section~\ref{related} for a complete description), the one closest in spirit to the IDE discussed in this paper is the online Dafny IDE \cite{leino2014dafny}.  For most of the exercises discussed in this paper, Dafny could be used as well.  However, the key difference that manifests itself the most for the purposes of this paper is our system's usage of a VC generation system \cite{harton:2011} that underlies the integrated Web IDE.  Using the VCs and a supporting prover capable of revealing which VCs fail to prove, it is possible to determine why a proof was unsuccessful from givens in the context. However, unlike the Dafny approach, which is backed by Z3 \cite{leino2010dafny}, the IDE presented here cannot be used to provide counter examples when verification fails.  The integrated prover does not use the proof-by-refutation technique, thus requiring a different sort of debugging to take its place. For example, a user contrasts what goal needs to be proved from the givens, tries to understand which givens would be more useful in attempting to prove the goal, and then adjusts the code or assertions as needed.

The reasoning process using the RESOLVE Web IDE is quite similar to what might be employed by one using a Coq-style proof assistant, except that the proofs to be done are mostly `obvious' due to the simple nature of VCs arising from well-designed software \cite{kirschenbaum:2009}. This characteristic has allowed us thus far to forego the need for manual proof assistance for VCs.\footnote{A proof assistant such as Coq or Isabelle \cite{nipkow2002isabelle} is indeed necessary for proving the results in reusable mathematical units employed by the automated prover, but the focus of this paper is on code correctness and VCs, assuming that the necessary theorems have been established previously \cite{kulczycki:icsr2013}.}  

To illustrate how the IDE helps produce correct code based on realtime feedback, we begin with a simple example that involves only the use of the \texttt{Integer} datatype.  This is followed by two object-based erroneous code examples: one that is recursive, and another that is iterative. These are examples presented to students as part of a software engineering course at Clemson.  In all cases, we follow an iterative approach that eventually leads to the correct code or adequate annotations.  The discussion concludes with a non-trivial queue copy example code with invariants that students were asked to develop for an assignment using the iterative approach.  We note that the examples discussed in this paper are meant to give an idea of the iterative process.  Several more complex components are available at the Web IDE; even more can be created by logging in to the site.  We conclude the paper with a discussion of the most related work and a summary.

\section{Understanding Design by Contract Using the IDE}
In this and following sections, we provide several illustrative examples, each building in complexity, that demonstrate the iterative process we envision when using the Web IDE to develop provably correct code. All examples discussed have a shared emphasis on the debugging aspect: that is, each requires sufficient knowledge of design by contract to correctly identify and amend a variety of errors in code or annotations based on interactive feedback from the prover in the form of VCs.

The first, relatively simple example presents an operation that arithmetically swaps the values of two \texttt{Integer} objects. Taking advantage of the conceptual simplicity of the code comprising this initial example, we also use this as an opportunity to familiarize readers with RESOLVE style specifications, syntax, and layout of the Web IDE. More details on the design of the RESOLVE language and its IDE may be found elsewhere \cite{cook:2014,sitaraman:2011,sitaraman:1994}.

\begin{figure}[htb] 
\centering
\frame{\includegraphics[width = \linewidth]{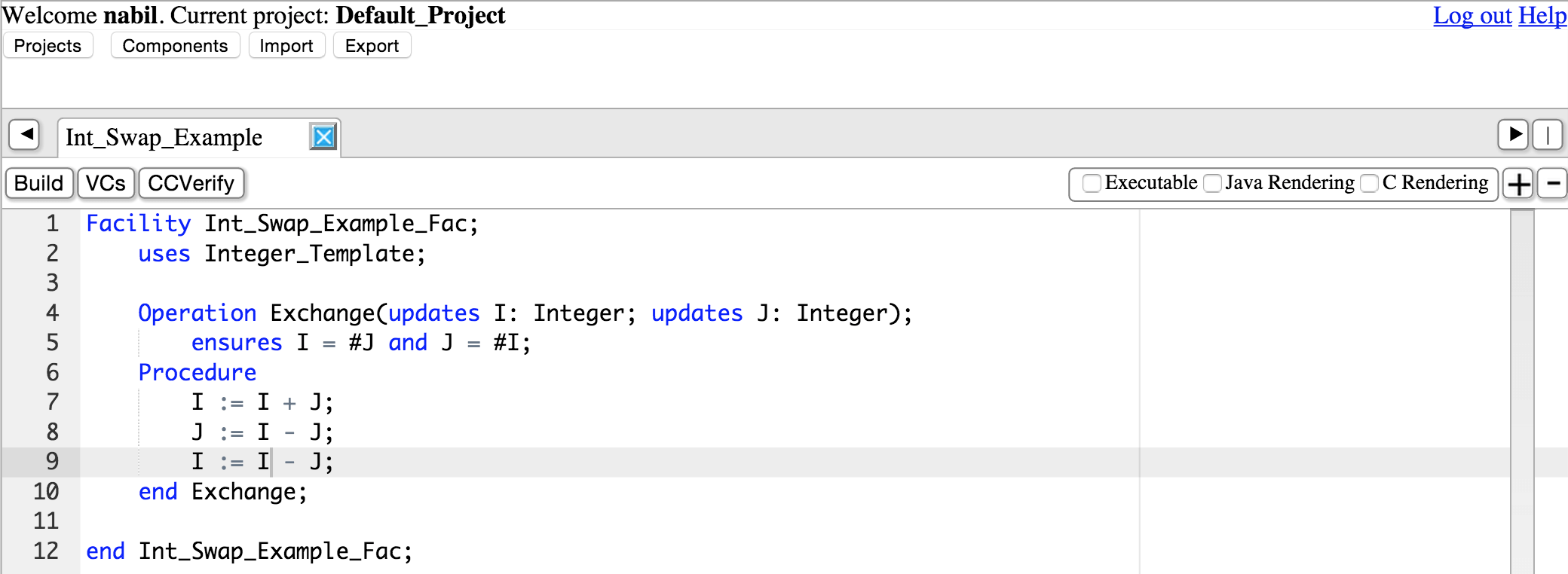}}
\caption[]{\texttt{Exchange} operation with missing requires clause.} 
\label{fig:intexchNoVC}
\end{figure}

Upon opening our first example, \texttt{Int\_Swap\_Example\_Fac} (Fig.~\ref{fig:intexchNoVC}), students are presented with code for a single operation, \texttt{Exchange}, that takes two \texttt{Integer} objects, denoted \texttt{I} and \texttt{J}. The essence of the specifications that formally communicate what exactly \texttt{Exchange} does can be found in the \texttt{ensures} clause (the postcondition), where we formally assert that $\texttt{I = \#J} \wedge \texttt{J = \#I}$. This assertion, when stated informally, can simply be read as ``the outgoing value of \texttt{I} is equal to the incoming value of \texttt{J} and the outgoing value of \texttt{J} is equal to the incoming value of \texttt{I}.''\footnote{In RESOLVE ensures clauses, \texttt{\#} denotes the \textit{incoming} value of a formal parameter.} Notice also that there is no return value for the \texttt{Exchange} operation. Instead, we prefix each parameter with mode \texttt{updates} to indicate to clients that each of the \texttt{Integer} values passed will contain a purposeful value as specified by the conclusion of the operation. 

Software developers are free to edit both the specifications (formal contracts) of \texttt{Exchange}, as well as its executable body (sandwiched between the \texttt{Procedure} and \texttt{end} keywords). When ready to verify the operation, students may invoke an integrated prover.  The exact prover used is of less importance to the discussion in this paper. It's worth noting here, however, that our system is supportive of three approaches: one based on term-rewriting (accessible via the \textit{RWVerify} button)  \cite{smith:2013}, another that is currently under development and uses a congruence closure algorithm in conjunction with a matcher for quantified expressions (accessible via \textit{CCVerify} button), and (optionally) an external SMT solver.\footnote{Z3 \cite{de2008z3} is currently being incorporated as a proving option.} The second one that is designed to be just sufficient to prove VCs arising from program correctness (as opposed to arbitrary mathematical assertions) is summarized in section \ref{sec:prover}, and that is the one used for the examples in this paper.

\begin{figure}[htb] 
\centering
\frame{\includegraphics[width = \linewidth]{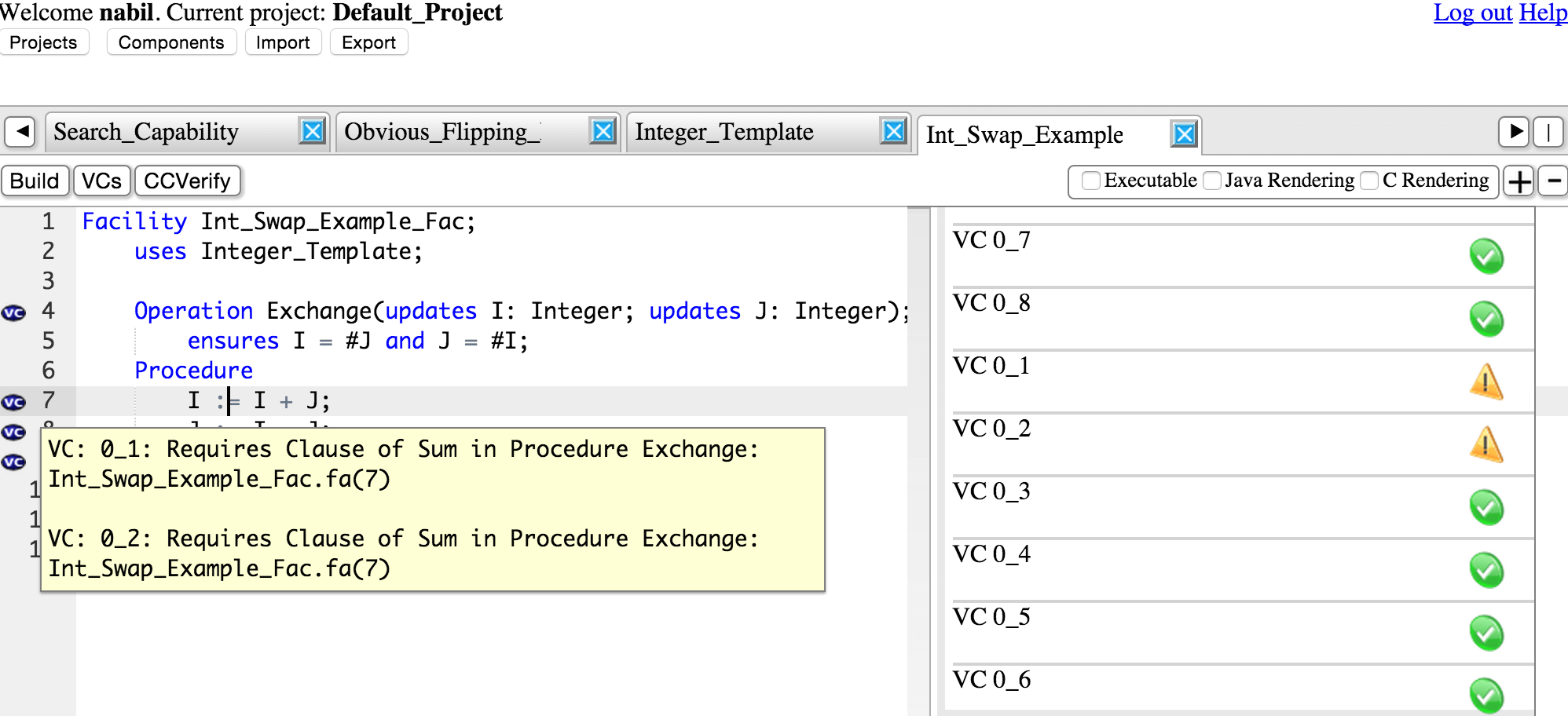}}
\caption[]{Proof attempt of \texttt{Exchange} operation with missing requires clause.} 
\label{fig:intexch0}
\end{figure}

Upon attempting to verify the \texttt{Exchange} operation, students are presented with a screen summarizing proof results, as shown in Fig.~\ref{fig:intexch0}. The system generates eight distinct VCs \cite{harton:2011}. VCs are mathematical assertions that are both necessary and sufficient for the code to be proven correct. To understand why there are eight VCs, we briefly describe the design-by-contract idea in this setting.  Two VCs arise from the two conjuncts of the \texttt{ensures} clause of the \texttt{Exchange} operation, guarantees to be provided by the implementer of the code.  Six VCs, two each for the \texttt{requires} clauses (preconditions) of each of the three statements in the code, namely that the sum or differences do not go outside computational integer bounds (i.e., \texttt{min\_int} and \texttt{max\_int}), for a total of eight VCs.  This is because preconditions of called operations are the responsibility of the calling code in design-by-contract.  Placing the cursor near the line number of a statement causes a box to appear referring to the names of one or more VCs generated if the statement produces VCs.

\begin{figure}[htb] 
\centering
\frame{\includegraphics[width = \linewidth]{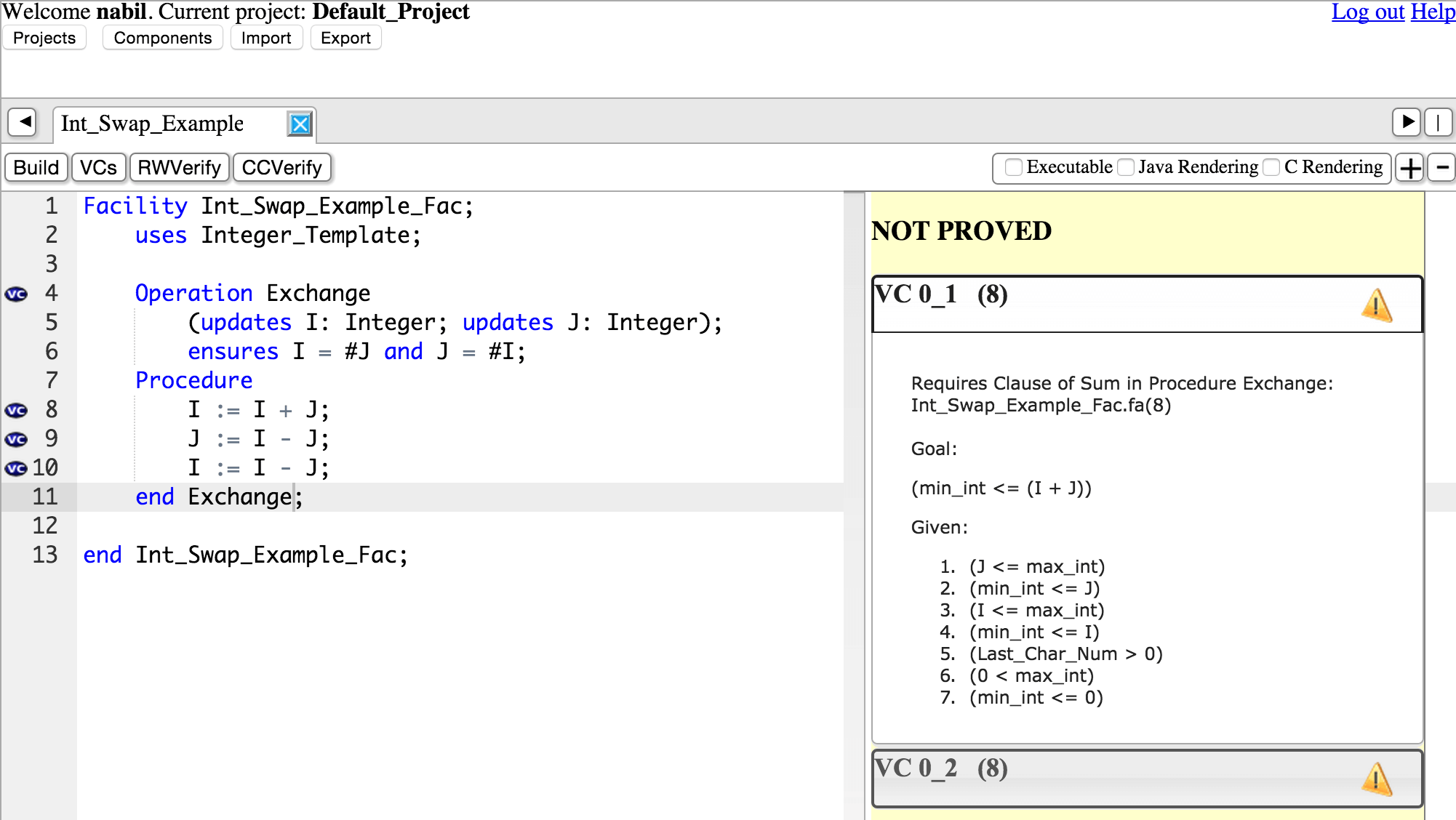}}
\caption[]{Full display of a VC in the IDE.} \label{fig:intexch1}
\end{figure}

Of the eight VCs, two are unable to be proven, as indicated by the yellow exclamation marks beside VC\_01 and VC\_02 (Fig.~\ref{fig:intexch1}).  The line numbers in code corresponding to the VC are given in parentheses.  While VCs in general might arise from any number of sources within a block of executable code, those unable to be established here arise from the requires clause of the sum operation that is implicitly invoked when \texttt{I} and \texttt{J} are added via the \texttt{+} operator. We leave it to a reader to convince themselves that an overflow or an underflow can occur in this code only for the first statement.   

To aid students in arriving at the particular insight necessary to debug this code, we encourage them to interactively explore the unprovable VCs by mousing over context sensitive VC buttons appearing next to lines of code that generated VCs. Upon clicking any of these buttons, the panel on the right hand side of the Web IDE updates with relevant, finer grained information about the particular VC queried, including a succinct description of what must be established (the goal) and the various facts (givens) the system currently knows.\footnote{A parsimonious approach to the generation of givens is under research and several of the unrelated givens are expected to disappear in the next version of the IDE.} 

\begin{figure}[htb]
\centering
\frame{\includegraphics[width = \linewidth]{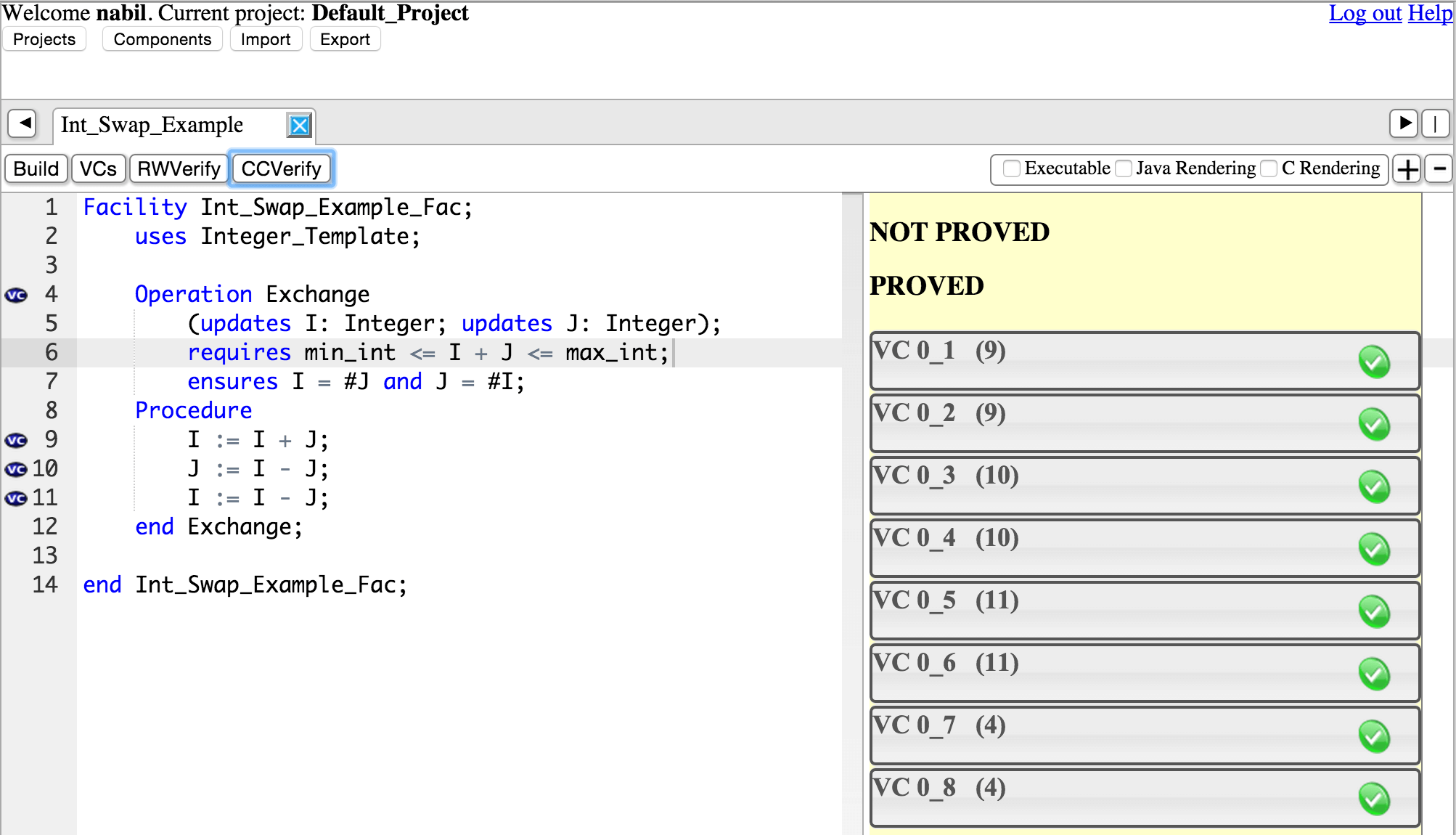}}
\caption[]{\texttt{Int\_Swap\_Example\_Fac} verified.}\label{fig:intexch2}
\end{figure}

In terms of the \texttt{Exchange} example, it is easy to observe that the system is unable to infer from the givens that \texttt{min\_int \textless=  (I + J)} (VC 0\_1) and 
\texttt{(I + J) \textless= max\_int} (VC 0\_2). It then becomes possible to infer that the system currently lacks knowledge suitable to realize that \texttt{Integer} overflow (or underflow) will not occur when the \texttt{+} operation is carried out. To remedy this, and `provide' the system with the assurance that this will not happen, students must defer to their knowledge of design-by-contract, amending the specification of \texttt{Exchange} with a suitable requires clauses as shown in Fig.~\ref{fig:intexch2}. Again, under design-by-contract, the requirements become givens to be used in proofs.  The figure shows that the Web IDE successfully verifies the code using the improved operation specification.

\section{Debugging Recursive Code}
\begin{figure}[htb]
\centering
\frame{\includegraphics[width=\linewidth]{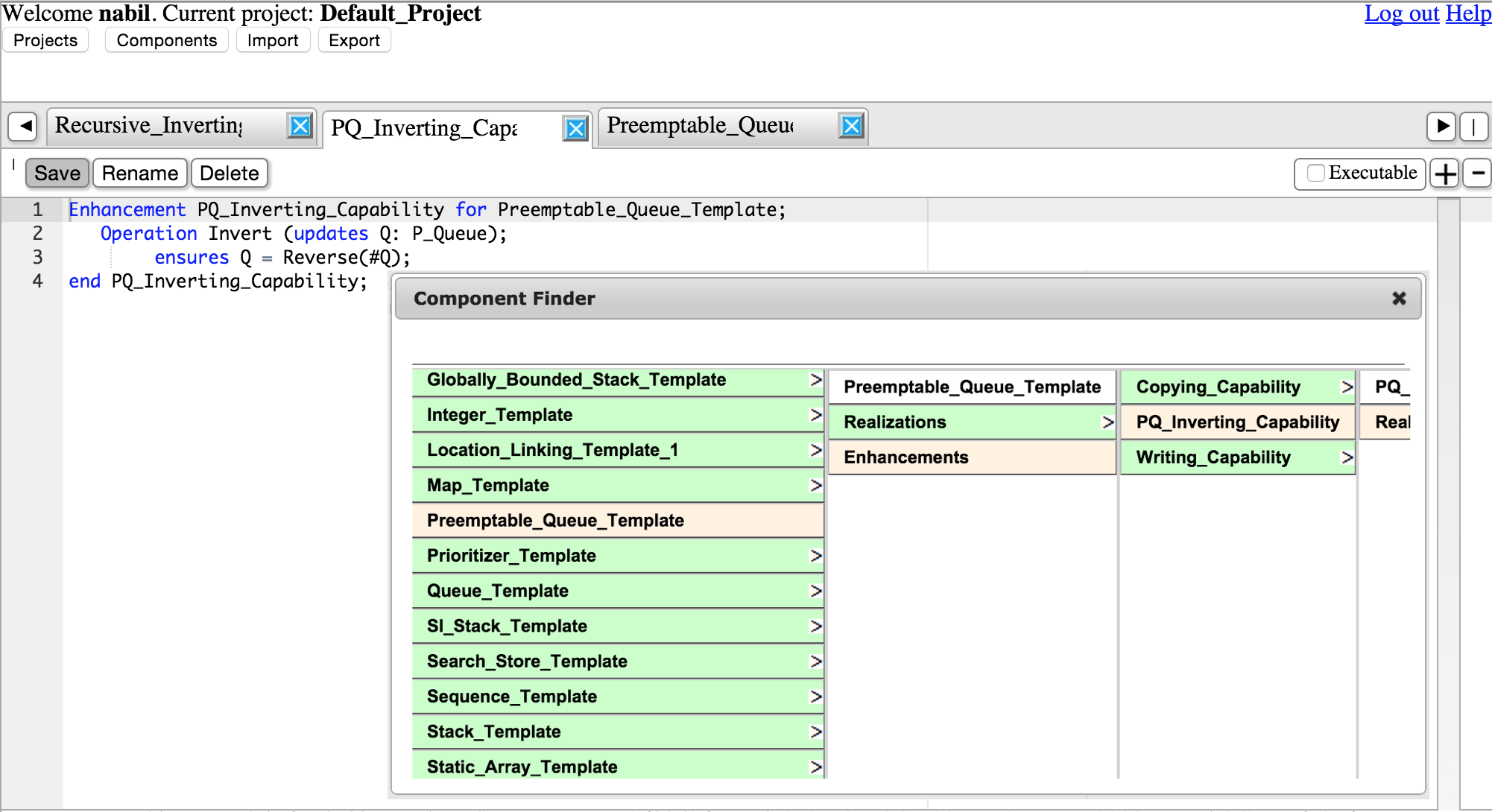}}
\caption[]{Selection of an enhancement in the Web IDE.}
\label{fig:enhancement}
\end{figure}

For the second example, we consider an example operation which inverts the order of items in a queue (see Fig.~\ref{fig:enhancement}). The \texttt{Invert} operation is specified in an enhancement (an extension using specification inheritance) to the \texttt{Preemptable\_Queue} concept and implemented in an enhancement realization, using only operations provided in the \texttt{Preemptable\_Queue} concept. This separation of concerns makes it possible to verify the enhancement realization in a modular fashion without referring to or refining to any one implementation of \texttt{Preemptable\_Queue} concept.  A preemptable queue differs from a regular queue in that it has operations to ``inject'' new items at the front of the queue (i.e., preempt the regular queue order), in addition to regular queue operations, such as \texttt{Enqueue} and \texttt{Dequeue}.  

\begin{figure}[htb]
\centering
\frame{\includegraphics[width=\linewidth]{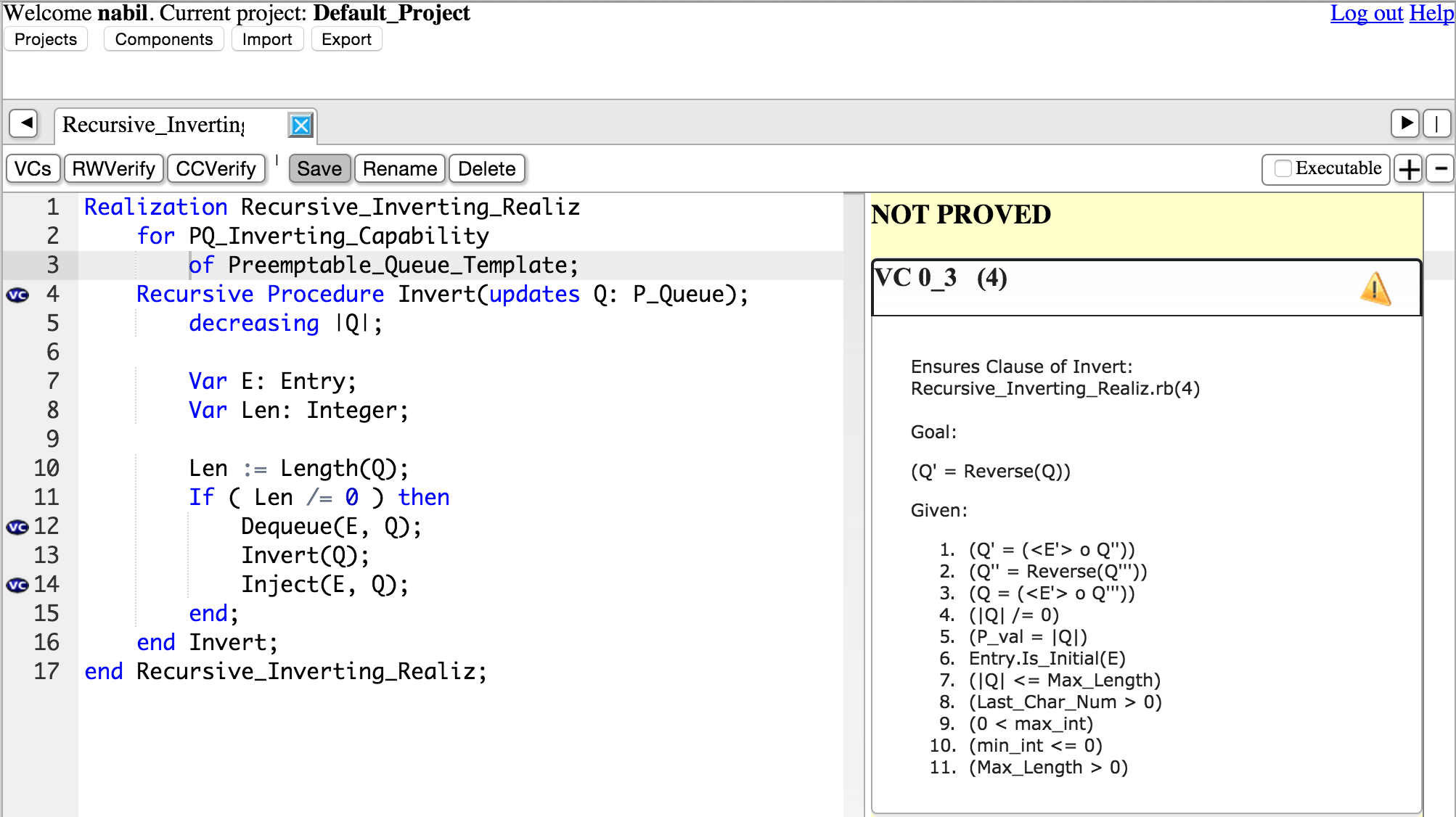}}
\caption[]{Inverting Code with Error.}
\label{fig:pq1}
\end{figure}

A complete specification of the \texttt{Preemptable\_Queue} concept is shown in Appendix \ref{App:PreQ}.  In the \texttt{Preemptable\_Queue} concept, the contents of the queue are conceptualized as a mathematical string (a structure similar to but simpler than a sequence in \texttt{Z}, because no positions are involved). So for this operation, the ensures clause (or post-condition) states that the outgoing value of the parameter \texttt{Q} should be the mathematical reverse of the input parameter (denoted by \texttt{\#Q}). Suppose that this operation is implemented using faulty code such as is shown in Fig.~\ref{fig:pq1}. Three of the VCs are verified, but VC 0\_3 is not. So as we did before, we encourage students to take a close look at that particular unprovable VC.

In the goal, \texttt{E'} is the dequeued entry, \texttt{Q'} is the conceptual string that stands for the value of the queue passed into the recursive call of \texttt{Invert}, and \texttt{Q} stands for the value of the queue at the beginning of the procedure. The goal is that \texttt{E'} concatenated with the reverse of \texttt{Q'} is equal to the reverse of \texttt{Q}. In order to debug this VC a user may first write down the goal and then apply transformations until we can clearly observe why the goal is unprovable.  The purpose here is to show a general process when the problem with the unprovable VC is less obvious.\footnote{While we show the details of these steps here, in actual debugging, such a detailed analysis may not be necessary; understanding of such principles as string concatenation is not commutative is straightforward and the problem may be inferred more readily.}

\begin{verbatim}
Goal: Q' = Reverse(Q)
\end{verbatim}

Our first transformation will be to use given \#1 and apply it to the left-hand side of the goal. We will label this transformation Step 1.

\begin{verbatim}
Step 1: <E'> o Q'' = Reverse(Q)
\end{verbatim}

Next, we will apply given \#2 to transform the left-hand side once again:

\begin{verbatim}
Step 2: <E'> o Reverse(Q''') = Reverse(Q)
\end{verbatim}

And then we apply given \#3 to the right-hand side:

\begin{verbatim}
Step 3: <E'> o Reverse(Q''') = Reverse(<E'> o Q''')
\end{verbatim}

Next, one can attempt to use a theorem from \texttt{String\_Theory}, which defines string notations and results involving those notations for mathematical strings. The theorem we need here states the following:

\begin{verbatim}
For all u, v : String, Reverse (u o v) = Reverse(v) o Reverse(u)
\end{verbatim}

This transformation will produce the following result:

\begin{verbatim}
Step 4: <E'> o Reverse(Q''') = Reverse(Q''') o Reverse(<E'>)
\end{verbatim}

Finally, we apply a theorem that states that the reverse of a single-length string is itself, which gives us step 5:

\begin{verbatim}
Step 4: <E'> o Reverse(Q''') = Reverse(Q''') o <E'>
\end{verbatim}

At this point, it is obvious to see that the goal is categorically false, as the concatenation operator is not commutative. Thus, the problem with the code is that the call to Inject is placing the dequeued entry on the wrong side of the recursively inverted queue. This example illustrates how the VC can serve as a guide to pinpoint the source of the error in formal reasoning.

In the case, the correction is to fix the code:  Specifically, the call to \texttt{Inject} needs to be replaced with a call to \texttt{Enqueue}.
\section{Loop Invariants}

This section outlines creation and iterative development of loop invariants for code using object-based examples.  Stacks and queues are abstract data types represented as objects in RESOLVE. Their behavior is specified in a \emph{Concept}, which is an abstract description of the methods all implementations must contain. It concludes with a discussion of an assignment given to students in a junior level software engineering course.
\subsection{Learning Iterative Invariant Development}
We begin with a simple example involving stacks to highlight the iterative steps we commonly see students working through with our Web IDE in reasoning about, and ultimately arriving at, appropriate assertions for loop invariants. Stacks, like queues, are modeled mathematically using strings; operations such as \texttt{Push} and \texttt{Pop} are specified using string notations.  

\begin{figure}[htb]
\centering
\frame{\includegraphics[width=\linewidth]{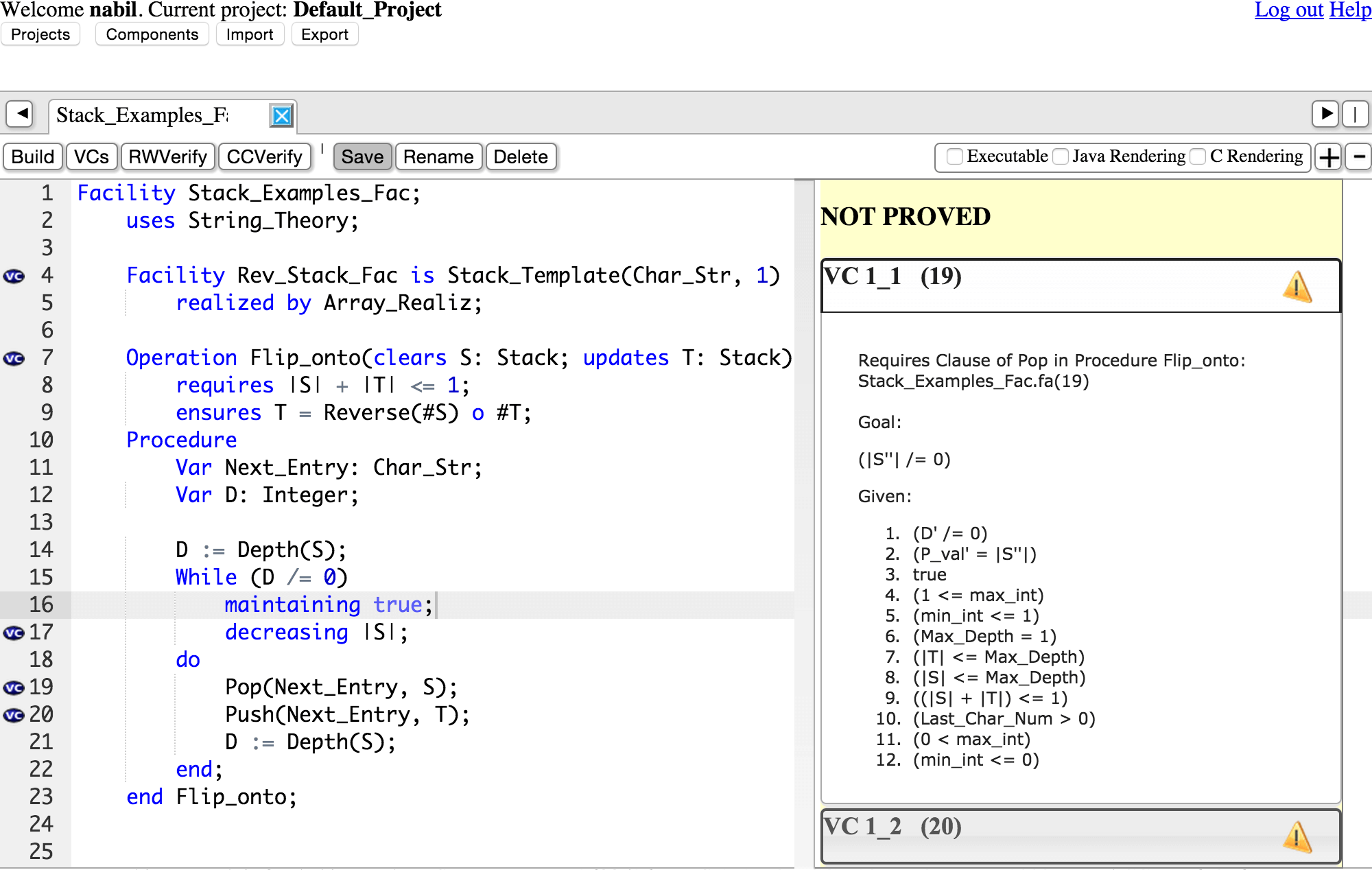}}
\caption[]{Loop with insufficient invariant.}\label{fig:stack1}
\end{figure}

Fig.~\ref{fig:stack1} shows an example operation presented in a classroom to teach the idea of invariants. \path{Flip_onto} iteratively transfers entries from a source stack, \texttt{S}, to a destination stack, \texttt{T}, resulting in a version of \texttt{T} that is prefixed by a `flipped' version of \texttt{S}. As expected, the intuition describing this outcome is formalized in the operation's \texttt{ensures} clause by the following succinct assertion: \texttt{T = Reverse(\#S) o \#T}. With the operation's input/output behavior formally expressed, students must turn to the task of deriving a suitable invariant for the while loop, expressed in RESOLVE using the \texttt{maintaining} clause. (The \texttt{decreasing} clause is used to document the progress metric necessary to prove termination.) 

Starting with a \texttt{maintaining} clause that simply reads ``\texttt{true}''---which is an appropriate starting point for beginners to understand the process of developing adequate invariants---the system (unsurprisingly) fails to establish correctness. Aside from the obvious inability to prove VCs corresponding to the operation's overall \texttt{ensures} clause, students using the Web IDE are able to see---with the help of the interactive VC buttons next to the line numbers---that VC 1\_1 and 1\_2 arising from calls to \texttt{Pop} and \texttt{Push} within the body of the loop are currently unprovable, as shown in Fig.~\ref{fig:stack1}. 

Examining VC 1\_1, students are immediately informed that the \texttt{requires} clause of \texttt{Pop} (\texttt{|S''| /= 0}) cannot be established. Referring to the list of available givens, students can see that while the system is aware that \texttt{D' /= 0}, it still lacks any knowledge relating the length of \texttt{S} to the current depth, \texttt{D}. To address this roadblock and provide the prover with the information it needs to meet the precondition criteria of \texttt{Pop}, students might start by amending the maintaining clause with the assertion that \texttt{|S| = D}. Sure enough, upon re-running the prover, students are given validation in the form of a green checkmark, indicating that one roadblock to verification of the current operation has been successfully dealt with.

In motivating further construction of the \texttt{maintaining} clause, students once again look to unproven VCs as a guide to development. In this case, looking specifically to VC 2\_2, students can see that the \texttt{ensures} clause to the overall operation is still unable to be established. Using this insight, combined with the goal this VC is attempting to establish---specifically, that \texttt{T' = (Reverse(S) o T)}---one way for students to proceed is to simply append this assertion to the evolving \texttt{maintaining} clause, yielding \texttt{|S| = D and T = (Reverse(\#S) o \#T)}. Upon doing so, students can indeed see that the prover is now able to establish the ensures clause of the operation (indicated by VC 2\_3), but is unable to establish the VC corresponding to the invariant of the while statement---suggesting that something is still lacking from the assertion. However, in examining the (now provable) goal of the overall \texttt{ensures} clause addressed in the previous step, students can see that it reads as follows: \texttt{(Reverse(S) o T) = (Reverse(S) o T)}. Thus, mirroring the same technique and intuition employed to make the ensures clause provable earlier---that is, adding \texttt{T = (Reverse(\#S) o \#T)}---students can now make the necessary cognitive leap to realize the clause must be changed to read: \texttt{Reverse(S) o T = Reverse(\#S) o \#T}, resulting in a final, provable assertion that reads: \texttt{D = |S| and Reverse(S) o T = Reverse(\#S) o \#T}.

\subsection{Applying Iterative Invariant Development}
Following an introduction to the iterative development of loop invariants and discussion, students used the Web IDE to complete reasoning assignments.  The assignments required students to write verified code for pre-specified concepts and enhancement operations.  The specification of one such operation to copy a generic \texttt{Preemptable\_Queue} is given below.\\
\\
\noindent\texttt{\textbf{\color{blue}Operation} Copy\_Queue \\
\indent(\textbf{\color{blue}restores} Q: P\_Queue; \textbf{\color{blue}replaces} Q\_Copy: P\_Queue);\\
\indent\textbf{\color{blue}ensures} Q\_Copy = Q; }\\

Table~\ref{table:assignment} is a summary of student performance for each of the invariant writing assignments. In addition to copying a queue, students wrote code for outputting a queue, reversing a sequence, and an end user application assignment that involved use of custom-made mathematical definitions and operations involving non-trivial types. The definitions in the end user assignment were not complemented by necessary results and hence, the prover was not of use in establishing the invariants.  The complexity of the assignment, the mathematics involved, and the absence of prover support are among possible reasons for the low success rate of students in developing appropriate invariants for those operations. 

\begin{table}[h]
\centering
\resizebox{\textwidth}{!}{%
\begin{tabular}{@{}lllll@{}}
\toprule
                                & \textbf{Writing (Queue)} & \textbf{Copying (Queue)} & \textbf{Reversal (Sequence)} & \textbf{Facility Operations} \\ \midrule
\textbf{Correct}                     & 70\%                     & 90\%                     & 60\%                         & 30\%                         \\
\textbf{Insufficient Invariant} & 20\%                     & 10\%                     & 30\%                         &                              \\
\textbf{Other}                        & 10\%                     &                              & 10\%                         & 70\%                         \\ \bottomrule
\end{tabular}
}\caption{Evaluation of Invariant Assignments}\label{table:assignment}
\end{table}

\begin{figure}[htb]
\centering
\frame{\includegraphics[width=\linewidth]{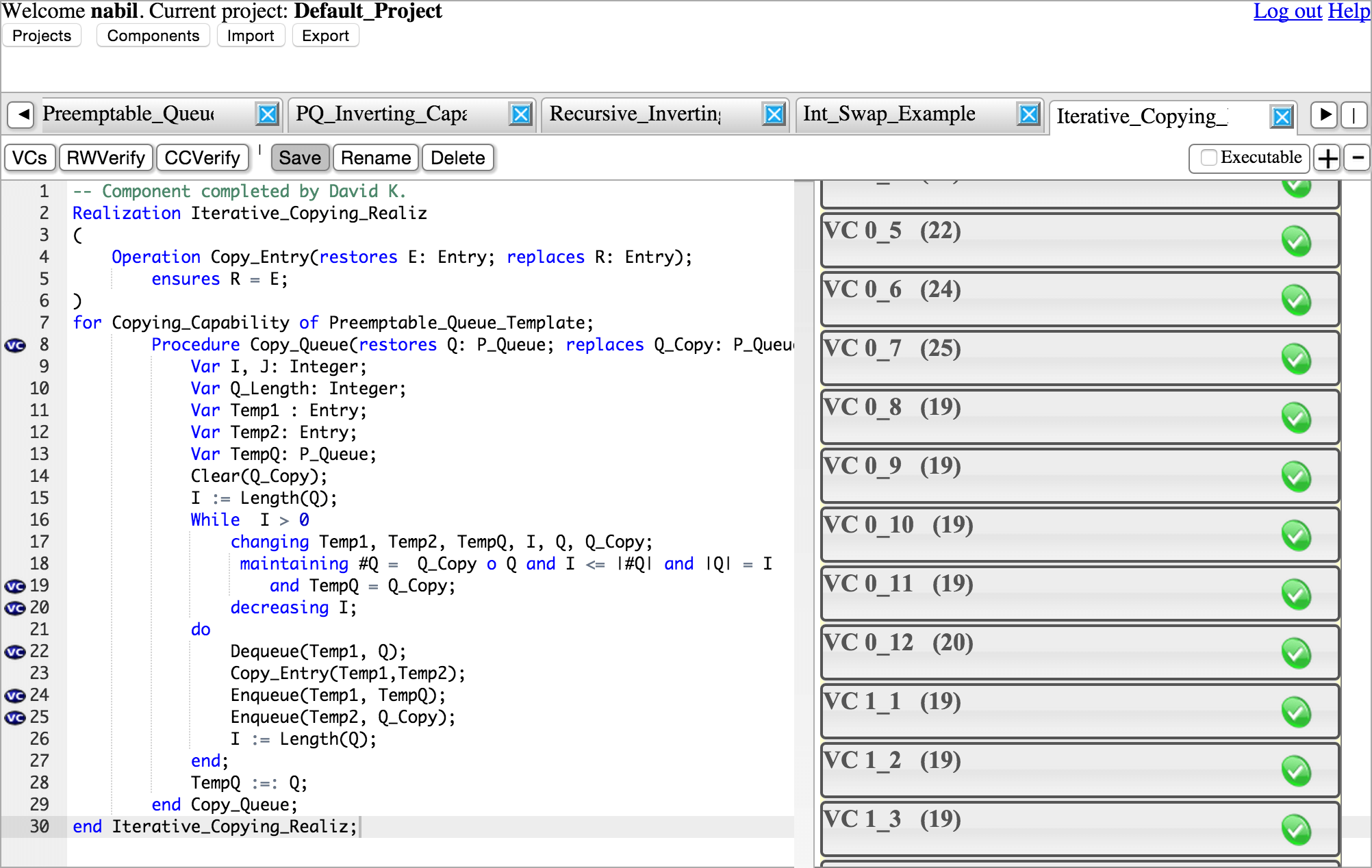}}
\caption[]{Student code for the Queue copy assignment.} \label{fig:studentcode}
\end{figure}

Fig.~\ref{fig:studentcode} is an example of code developed by a student for the \texttt{Copy\_Queue} assignment.\footnote{In the figure, \texttt{changing} clause is optional and it lists variables potentially affected by the loop; variables not mentioned are assumed to be unchanging.  In the absence of this clause, all variables are assumed to be affected.  This clause is useful to simplify some routine invariants \cite{harton:2011}.}  Neither the code nor the invariant is necessarily optimal.  Proofs of all 18 VCs generated for this copy operation are completed in an average time of 3 seconds (total) on the server that hosts the Web IDE.  As noted earlier in the introduction, other provers, such as Z3, could be used to discharge the VCs.

\section{Summary of the Prover Underlying the Web IDE} \label{sec:prover}
The verifying compiler that serves as the back end of the web interface contains a modular VC generation subsystem \cite{cook2012specification,harton:2011} which provides input to an automated  prover.   The automated prover relies on previously proven results in a library of mathematical theories that are reused in the specification of programming concepts \cite{smith:2013}.  

At the core of the \emph{CCVerify} automated VC prover is a congruence closure algorithm that incorporates the Theory of Equality, similar in spirit to that described in \cite{Fast}. An outer layer that incorporates pattern matching techniques for expressions containing universally quantified variables is engaged, similar to the matcher described in \cite{Simplify}. In this way, a single component can handle problems from multiple domains. \emph{CCVerify}, which contains fewer than 2000 lines of Java code, is designed to be fast, simple, and effective. As it is fully integrated into the compiler, there are no issues with portability, licensing, or translation of the assumptions to other formats as there might be if an external tool were used.  

There is an implied division of labor in the production of proofs. Specifications must be written so that the consequent of the VC eventually produced is  a predicate with constants as arguments. It turns out that typically it is sufficient to use only instances of previously proven universally quantified statements (these are members of a reusable theorem library) to construct a proof, assuming that the specifications used to generate the VC make such a proof relatively obvious. Sophisticated techniques used in automated theorem provers may not be required to discharge the VCs.  We are currently testing this hypothesis using a limited version of Z3 as well.

Our mathematical specification system is feature--rich.  It allows for polymorphic types and first class functions. These features make a \emph{direct} translation of \emph{some} of our mathematical theories to the standard many-sorted first-order logic language \cite{barrett2010smt} used in SMT proving  impossible, though it is possible to support these features relatively simple (in a sound, but not complete way) within the matching component of the integrated prover.  

\section{Related Work}\label{related}
A summary of related specification/verification languages may be found in \cite{hatcliff2012behavioral}, and tools or IDEs to facilitate their usage are discussed in the first proceedings of this workshop \cite{DBLP:journals/corr/DuboisGM14}. We discuss only the most related work in this section.

Like RESOLVE that underlies the Web IDE discussed in this paper, Dafny is a programming and specification language intended for verification of functional correctness \cite{leino2014dafny}.  The Eiffel integrated language effort \cite{EiffelRef}, though initially focused on runtime assertion checking, is now supported by Eve~\cite{furiaeptcs}, the Eiffel Verification Environment, which includes the AutoProof~\cite{DBLP:journals/corr/abs-1106-4700} tool for static verification.  Both Dafny and AutoProof translate code and specifications into Boogie~\cite{DBLP:conf/fmco/BarnettCDJL05}, an intermediate verification language.  The Boogie tool can create VCs suitable as input to an SMT solver.  An important distinction is that the RESOLVE compiler handles VC generation internally, and displays them in a format that makes it easy for users to connect problematic VCs with the code and specifications that produced them.

Java Modeling Language (JML) is a specification/verification language for Java programs, and tools for the language include the ability to do runtime assertion checking \cite{leavens2006preliminary}. JML does not have an IDE, but there are efforts to integrate JML as plugin to Eclipse \cite{chalin2008jml4,cok2014openjml}. Tools are available for ProB, an animation and a model checker for the B-Method, which is a formal method based on abstract 
machine notation \cite{bendisposto2014prob,witulski2014prob}. The VeriFast effort is aimed at verifying single/multi-threaded C and Java programs \cite{jacobs2011verifast,smans2013verifast}. VeriFast also includes a GUI that is packaged into their code distribution. 

We are not alone in employing a formal methods IDE in education.  Whereas our educational focus is mostly on software engineering aspects (though we have used the Web IDE to a limited extent in a discrete mathematics course), teaching discrete mathematics and specifications using an IDE is the focus of FoCaLiZe---an IDE 
that takes source code, specification properties, and machine-checkable proofs to produce executable OCaml 
code and checkable Coq input values \cite{jaume2014teaching,pessaux2014focalize}.  Though our Web IDE does not support inferring loop invariants, invariant inference is a useful feature; an Eclipse plugin with a goal to infer object and loop invariants for C programs is discussed in \cite{cok2014speedy}.  

\section{Conclusions}
This paper has detailed an iterative approach for creating, debugging, and developing components that are correct with respect to their specifications, using an IDE equipped with a verification system.  Using several illustrative examples drawn from lectures and student assigments, we have explained how students and software engineers, in general, can develop provably correct software iteratively based on the VCs and feedback received from the RESOLVE Web IDE.  Extensive experience with the IDE in the classroom indicates that students are indeed capable of producing correct software using the the IDE as discussed in this paper.  While the present paper has focused only on functional correctness of code, the IDE includes features to create and view mathematical units and data abstraction realizations with representation invariants and abstraction relations, as well as for generating executable Java code from RESOLVE code \cite{welch:2014}. 

A variety of improvements to the IDE are in progress, ranging from minor visual improvements, such as highlighting VC buttons that correspond to unprovable VCs, to more significant ones, such as the creation and development of performance specifications and related correctness checks.

\section{Acknowledgements}
The RESOLVE verifying compiler is a multi-decade project involving researchers at several universities, including, but not limited to, Clemson University, Ohio State University, and Denison University.  We acknowledge the ideas and support of members of the group in this endeavor. This research has been funded in part by the US NSF grants CCF-0811748, CCF-1161916, and DUE-1022941.

\bibliography{references}
\newpage
\appendix
\section{Preemptable\_Queue Specification} \label{App:PreQ}
\begin{figure}[h]
\centering
\frame{\includegraphics[width=\linewidth]{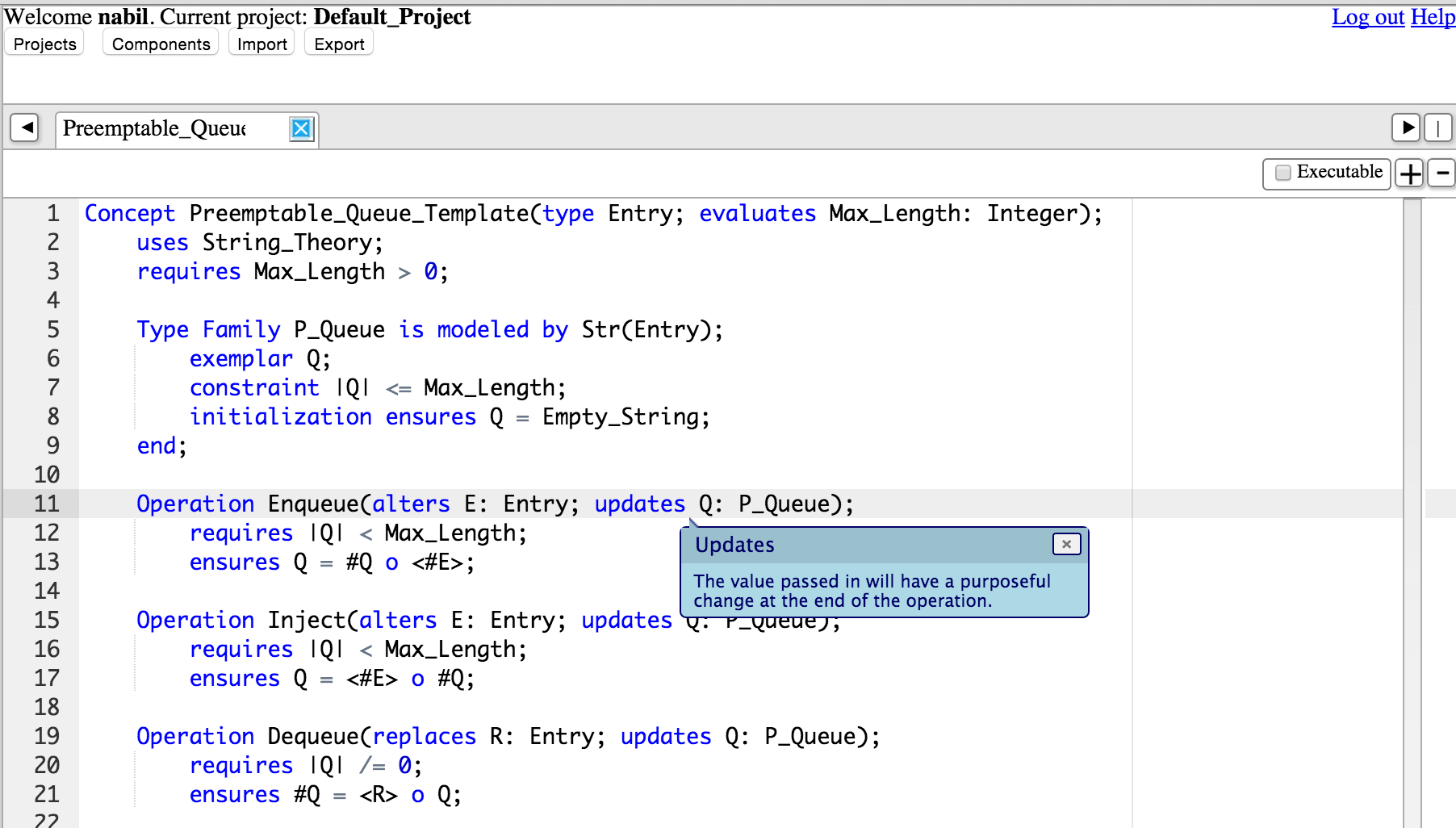}}
\caption[]{Partial \texttt{Preemptable\_Queue} specification.}\label{fig:spec}
\end{figure}

\end{document}